\documentclass[a4paper,11pt]{article}
\usepackage{jheppub}
\usepackage{subcaption}
\DeclareUnicodeCharacter{202F}{\,}

\usepackage{xcolor}
\usepackage{float}
\definecolor{llblue}{HTML}{07839C}
\definecolor{ljgreen}{HTML}{006400}

\title{\boldmath Global analysis of the $\rho$ resonance in lattice QCD}

\author[a]{Luka Jevsenak,}
\author[a,b]{Luka Leskovec,}
\author[c]{Stefan Meinel,}
\author[e]{John Negele,}
\author[f]{Srijit Paul}
\author[e]{and Andrew Pochinsky}

\affiliation[a]{Faculty of Mathematics and Physics, University of Ljubljana, Jadranska 19, 1000 Ljubljana, Slovenia}
\affiliation[b]{Jo\v zef Stefan Institute, Jamova 39, 1000 Ljubljana, Slovenia}

\affiliation[c]{Department of Physics, University of Arizona, Tucson, AZ 85721, USA}

\affiliation[e]{Center for Theoretical Physics, Massachusetts Institute of Technology, Cambridge, MA 02139, USA}

\affiliation[f]{Department of Physics, University of Cyprus, Aglantzia 2109, Nicosia, Cyprus}

\emailAdd{luka.jevsenak@fmf.uni-lj.si}
\emailAdd{luka.leskovec@ijs.si}
\emailAdd{smeinel@arizona.edu}

\abstract{
We present a lattice-QCD study of the $\rho$ resonance in which we address the quark-mass and lattice-spacing dependence in a novel way using a global analysis of four $N_f=2+1$ ensembles. We parametrize discretization effects in the lattice energy levels and not in the scattering amplitude, which is defined in the continuum. We describe the amplitude using a $K$-matrix parameterization, which contains the light- and strange-quark mass dependence. By varying the parameterizations of quark-mass and lattice-spacing dependence, we examine the sensitivity of the physical-limit $\rho$ pole position to these choices and obtain the result $\sqrt{s}_\rho = 744(11)(7) - i\,69(7)(3)~\text{MeV}$ from a model average.

}

\begin{document}
\maketitle
\flushbottom

%%%%%%%%%%%%%%%%%%%%%%%%%%%%%%%%%%%%%%%%%%%%%%%%%%%%%%%%%%%%%%%%%%%%%%%%%%%%%%%%%%%%%%%%%%%%%%%%%%%%%%%%%%%%%%%%%%%%%%%%%%%%%%%%%%%%%%%%%%%%%%%%%%%%%%%%%%%%%%%%%%%%%%%%%%%%%%%%%%%%%%%%%%%%%%%%%%%%%%%%%%%%%%%%%%%%%%%%
\section{Introduction}
The $\rho$ resonance is often considered as a benchmark for lattice QCD calculations of hadron-hadron scattering. It appears in $I=1$ $P$-wave $\pi\pi$ scattering and is considered a test of lattice methods, ranging from correlation-function methodology through spectrum extraction to finite-volume analysis. Past calculations have established the extraction of its phase shift and resonance parameters at various quark masses and lattice spacings, improved the resolution of the amplitude, and extended the analysis towards physical quark masses~\cite{CP-PACS:2007wro, Feng:2010es, Lang:2011mn, Dudek:2012xn, Wilson:2015dqa, Bulava:2016mks, Alexandrou:2017mpi, Bali:2015gji, Fischer:2020yvw, Boyle:2024grr}. With the plethora of work done, the lattice community has established methods and workflows that bring us closer to understanding strong interactions \cite{Briceno:2017max}. Precision calculations are a crucial component of this effort, and the $\rho$ is a natural target for precision resonance physics.

Alongside our goal of developing lattice methods and understanding the strong interaction comes an important motivation for a precise understanding of resonances; resonances appear in many electroweak processes that we use to determine Standard Model parameters. One such example is $B\to\pi\pi\ell\bar\nu$ \cite{Leskovec:2025gsw}, from which we can determine $|V_{ub}|$; there, the underlying $\pi\pi$ scattering amplitude enters both the relation between finite-volume matrix elements \cite{Lellouch:2000pv, Lin:2001ek, Briceno:2015csa, Briceno:2021xlc} and infinite-volume transition amplitudes and their continuation to the resonance pole. 

Connecting lattice QCD calculations to nature requires control over the quark-mass dependence and discretization effects in the scattering amplitude. Previous studies investigated these effects by determining resonance parameters separately on each ensemble and then extrapolating these parameters to the continuum limit (in some cases) and physical quark masses \cite{Andersen:2018mau, ExtendedTwistedMass:2019omo, Fischer:2020yvw, Boyle:2024grr, Wang:2025hew}. Different approaches were used in extrapolation in the light-quark mass, ranging from the inverse amplitude method (IAM) to unitarized chiral approaches, which relate the scattering amplitude to the quark-mass dependence of pseudoscalar masses and decay constants through a common effective field theory (EFT) description~\cite{Hu:2017wli, Fischer:2020yvw, Molina:2020qpw}. Refs.~\cite{Hu:2016shf, Hu:2017wli} investigated the role of the $K\bar K$ channel in comparing two- and three-flavor results. Within three-flavor QCD, ref.~\cite{Molina:2020qpw} analyzed the light- and strange-quark mass dependence along trajectories with fixed strange-quark mass or fixed sum of quark masses. This trajectory dependence is distinct from the absence of a dynamical strange quark in two-flavor calculations. It motivates treating the light- and strange-mass dependence together, so that changes in the strange-quark mass are not absorbed into the inferred light-quark-mass dependence.

In this work, we address the quark-mass and lattice-spacing dependence in a novel way using a global analysis of four $N_f=2+1$ ensembles. Unlike earlier work, we parametrize discretization effects in the lattice energy levels and not in the scattering amplitude, which is naturally defined only in the continuum. We describe the latter using a $K$-matrix parameterization, which contains the light- and strange-quark mass dependence. We perform a global fit to the energy spectra on all ensembles, where the model is given by the sum of the energy levels predicted for each choice of quark masses and lattice volume but at zero lattice spacing using the L\"uscher quantization condition \cite{Luscher:1990ux} and lattice-spacing dependent terms.
The quark-mass dependence is described by EFT-inspired functional forms. By varying the parameterizations of quark-mass and lattice-spacing dependence, we examine the sensitivity of the $\rho$ pole to these choices and combine the results through model averaging.

%%%%%%%%%%%%%%%%%%%%%%%%%%%%%%%%%%%%%%%%%%%%%%%%%%%%%%%%%%%%%%%%%%%%%%%%%%%%%%%%%%%%%%%%%%%%%%%%%%%%%%%%%%%%%%%%%%%%%%%%%%%%%%%%%%%%%%%%%%%%%%%%%%%%%%%%%%%%%%%%%%%%%%%%%%%%%%%%%%%%%%%%%%%%%%%%%%%%%%%%%%%%%%%%%%%%%%%%
\section{Gauge-field ensembles}

Our calculation is performed on four different gauge-field ensembles with $N_f = 2 + 1$ dynamical quark flavors, using a Symanzik gauge action\cite{Symanzik:1983pq, Symanzik:1983dc, Symanzik:1983gh, Luscher:1985zq} and a clover-improved Wilson action \cite{Wilson:1974sk, Sheikholeslami:1985ij}. Details of these actions and of the generation of the ensembles can be found in ref.~\cite{Yoo:2026uul}, although here we use our own determinations of the lattice spacings. 
The main properties of the four ensembles that we use are summarized in  Table \ref{tab:ens}. 
These four ensembles give us a minimal set in which we can fit and study the discretization effects and quark-mass dependence simultaneously.
The \texttt{C13} ensemble has the heaviest pion mass and largest lattice spacing, as stated in the Table \ref{tab:ens}. The \texttt{D5} and \texttt{D6} ensembles have similar lattice spacing, with \texttt{D6} at significantly lighter pion mass compared to the other three ensembles. Lastly, the \texttt{E5} ensemble has a pion mass similar to that of the \texttt{D5} and the smallest lattice spacing.

The lattice-spacing values for the \texttt{C13}, \texttt{D6}, and \texttt{E5} ensembles given in Table \ref{tab:ens} were determined using the $\Upsilon(2S)-\Upsilon(1S)$ splitting computed with the same lattice NRQCD action as the order-$v^4$ action in ref.~\cite{Meinel:2010pv}, except that the stability parameter was set to $n=4$. The bottom-quark masses were tuned in a self-consistent way using the spin-averaged $1S$ bottomonium kinetic mass. For the \texttt{D5} ensemble, the NRQCD calculation was not performed, and we instead use $a_{\texttt{D5}}=\frac{(w_0/a)_{\texttt{D6}}}{(w_0/a)_{\texttt{D5}}}a_{\texttt{D6}}$, where $a_{\texttt{D6}}$ is the lattice spacing of \texttt{D6} from the $\Upsilon(2S)-\Upsilon(1S)$ splitting, and $w_0/a$ denotes the Wilson-flow scale in lattice units as defined in ref.~\cite{BMW:2012hcm}.

\begin{table}[htbp]
\centering
\begin{tabular}{c|c|c|c|c}
\hline
& \texttt{C13} & \texttt{D5} & \texttt{D6} & \texttt{E5} \\
\hline
$N_s^3\times N_t$ & $32^3\times 96$ & $32^3\times 64$ & $48^3\times 96$ & $48^3\times 128$ \\
$\beta$ & $6.1$ & $6.3$ & $6.3$ & $6.5$ \\
$am_{u,d}$ & $-0.285$ & $–0.2390$ & $-0.2416$ & $–0.2070$ \\
$am_s$ & $-0.245$ & $–0.2050$ & $-0.205$ & $–0.1750$ \\
$C_{SW}$ & $1.2493$ & $1.2054$ & $1.2054$ & $1.1701$ \\
$a$ [fm] & $0.11403(77)$ & $0.08990(82)$ & $0.08766(79)$ & $0.06827 (55)$ \\
$L$ [fm] & $3.649(25)$ & $2.877(26)$ & $4.208(38)$ & $3.277(26)$ \\
$am_\pi$ & $0.18295(36)$ & $0.12885(44)$ & $0.07760(36)$ & $0.10030(23)$ \\
$am_K$ & $0.30475(17)$ & $0.24431(26)$ & $0.22809(16)$ & $0.18209(17)$ \\
$am_{\eta_s}$ & $0.3882(19)$ & $0.32020(15)$ & $0.31224(15)$ & $0.28423(39)$ \\
$m_\pi$ [MeV] & $316.6(2.2)$ & $282.8(2.8)$ & $174.7(1.8)$ & $289.9(2.4)$ \\
$m_K$ [MeV] & $527.4(3.6)$ & $536.3(4.9)$ & $513.4(4.6)$ & $526.3(4.3)$ \\
$m_{\eta_s}$ [MeV] & $671.8(5.6)$ & $702.8(6.4)$ & $702.8(6.3)$ & $821.5(6.7)$ \\
$N_{\rm config}$ & $1039$ & $467$ & $621$ & $297$ \\
$N_{\rm source}$ & $8$ & $16$ & $16$ & $16$ \\
\hline
\end{tabular}
\caption{Parameters of the gauge-field ensembles.\label{tab:ens}}
\end{table}

\section{Two-point correlation functions} \label{sec:2pt}
We construct the two-point correlation functions using a combination of single-meson and two-meson interpolators. The correlation functions are constructed in the same way as in \cite{Alexandrou:2017mpi}. We use two types of single meson interpolators:
\begin{align}
 O_{qq} &= O_1 = \left[\bar{q}\gamma_i q\right]\left(\mathbf{P}\right),\\
 O_{qq,t} &= O_2 = \left[\bar{q}\gamma_t\gamma_i q\right]\left(\mathbf{P}\right),
\end{align}
with total momentum $\mathbf{P}$. For the \texttt{C13} and \texttt{D5} ensembles, we use both $O_1$ and $O_2$ interpolators, and for the \texttt{D6} and \texttt{E5} ensembles, we only calculate the Wick contractions with the $O_1$ interpolator. 

The two-meson interpolators are all in the form
\begin{align}
 O_{mm,o} = O_{2+o} = \left[\bar{q}\gamma_5 q \right] \left( \mathbf{p}_{1,o} \right) \left[ \bar{q}\gamma_5 q \right] \left( \mathbf{p}_{2,o} \right),    
\end{align}
where $\mathbf{p}_{1,o}+\mathbf{p}_{2,o} = \mathbf{P}$, and the index $o$ runs over different momenta combinations.

The quark fields are Wuppertal-smeared~\cite{Gusken:1989ad} using APE-smeared links~\cite{APE:1987ehd} in the Laplacian. The smearing coefficients $\alpha_{\rm wup}$ and $\alpha_{\rm APE}$ and iteration counts $N_{\rm wup}$ and $N_{\rm APE}$ are listed in Table~\ref{tab:smearing}. There are four possible types of Wick contractions for these interpolators. All of them are described in \cite{Alexandrou:2017mpi}. We compute the contractions for all possible interpolator combinations with $\left|\mathbf{P}\right|^2, \left|\mathbf{p_1}\right|^2, \left|\mathbf{p_2}\right|^2 \leq 3 \left(\frac{2\pi}{L}\right)^2$.  

We project each interpolating field onto irreducible representations of the relevant finite-volume Little Groups, $O_i^{\mathbf{P},\Lambda,r}$, as in ref.~\cite{Alexandrou:2018jbt}. This gives two-point correlation functions with well defined symmetries:
\begin{align}
    C^{\mathbf{P},\Lambda,r}_{ij}(t) = \left<\left[O^{\mathbf{P},\Lambda,r}_i(t)\right]^\dagger O^{\mathbf{P},\Lambda,r}_j(0)\right>,
\end{align}
where $\Lambda$ labels the lattice irreducible representation and $r$ the row of the irreducible representation. The projected correlation functions that are equivalent by symmetry are averaged. These averaged correlation functions then still depend on the interpolator indices $i,j$, irreducible representation $\Lambda$, and size of total momentum $\hat{P} = |\mathbf{P}|^2 \left(\frac{2\pi}{L}\right)^{-2}$.  We average over all different directions of total momentum $\mathbf{P}$ and all rows $r$ of the same $\Lambda$. For example, the $\hat{P}=1, \Lambda = E$ correlation function is obtained by averaging over all correlation functions with $\mathbf{P} \in [001, 010, 100, 00-1, 0-10, -100]\left(\frac{2\pi}{L}\right),\ \Lambda = E,\ r \in \{1,2\}$.

The projected and averaged correlation functions form a correlation matrix for each combination of $\hat{P},\Lambda$: $C^{\hat{P},\Lambda}_{ij,h}(t)$. 
The indices $i,j$ run over all interpolators, where $i=1,2$ corresponds to the 1-2 single meson and $N$ two meson interpolators, with $N$ being dependent on the irreducible representation. All combinations of $\hat{P},\Lambda$ and their respective $N$ are listed in Table \ref{tab:operators}. To account for error propagation throughout the energy spectrum extraction (sections \ref{sec:2pt} and \ref{sec:extraction}), we use the jackknife method. Index $h$ is the jackknife sample index. We generate jackknife samples by averaging correlation functions over all sources within the same configuration and then resampling over the configurations:
\begin{align}
    C_h = \frac{1}{\left(N_{\rm config}-1\right)} \sum_{c=1; c\neq h}^{N_{\rm config}} \frac{1}{N_{\rm source}}  \sum_{s=1}^{N_{\rm source}} C_{c,s}.
\end{align}
The number of configurations and sources per configuration is listed in the Table \ref{tab:ens}.

\begin{table}[htbp]
\centering
\begin{tabular}{c|c|c|c|c|c|c|c|c}
\hline
$\hat{P}=\left|\mathbf{P}\right|^2\left(\frac{2\pi}{L}\right)^{-2}, \Lambda$ & $0,T_1$ & $1,A_2$ & $1,E$ & $2,B_1$ & $2,B_2$ & $2,B_3$ & $3,A_2$ & $3,E$ \\
\hline 
$N(O_{mm})$ & $3$ & $3$ & $2$ & $2$ & $2$ & $2$ & $2$ & $1$ \\ 
\hline
\end{tabular}
\caption{ Every irreducible representation we can project to from total momentum $\hat{P} \leq 3$ and angular momentum $l=1$. For each irreducible representation, we list the number of projected two-meson interpolators we can construct. \label{tab:operators}}
\end{table}

\begin{table}[htbp]
\centering
\begin{tabular}{c|c|c|c|c}
\hline
& \texttt{C13} & \texttt{D5} & \texttt{D6} & \texttt{E5} \\
\hline
$\alpha_{\rm wup}$   & $3.0$ & $3.0$ & $3.0$ & $3.0$ \\
$N_{\rm wup}$        & $20$  & $20$  & $55$  & $80$  \\
$\alpha_{\rm APE}$   & $2.5$ & $2.5$ & $2.5$ & $2.5$ \\
$N_{\rm APE}$        & $25$  & $25$  & $32$  & $25$  \\
\hline
\end{tabular}
\caption{Wuppertal smearing parameters for the quark fields and APE smearing parameters for the gauge links in the smearing kernel. \label{tab:smearing}}
\end{table}

%%%%%%%%%%%%%%%%%%%%%%%%%%%%%%%%%%%%%%%%%%%%%%%%%%%%%%%%%%%%%%%%%%%%%%%%%%%%%%%%%%%%%%%%%%%%%%%%%%%%%%%%%%%%%%%%%%%%%%%%%%%%%%%%%%%%%%%%%%%%%%%%%%%%%%%%%%%%%%%%%%%%%%%%%%%%%%%%%%%%%%%%%%%%%%%%%%%%%%%%%%%%%%%%%%%%%%%%
\section{Energy spectrum extraction} \label{sec:extraction}
To determine the finite-volume spectra in the irreducible representations listed in Table \ref{tab:operators}, we solve the generalized eigenvalue problem for each of the two-point correlation matrices to extract the principal correlators $\lambda^{\hat{P},\Lambda,n}$:
\begin{align}
    C^{\hat{P},\Lambda}_{ij}(t) v_j^{\hat{P},\Lambda,n}(t) = \lambda^{\hat{P},\Lambda,n} (t,t_0) C^{\hat{P},\Lambda}_{ij}(t_0) v_j^{\hat{P},\Lambda,n}(t).
\end{align}

\begin{figure}[htb!]
\centering
\includegraphics[width=\textwidth]{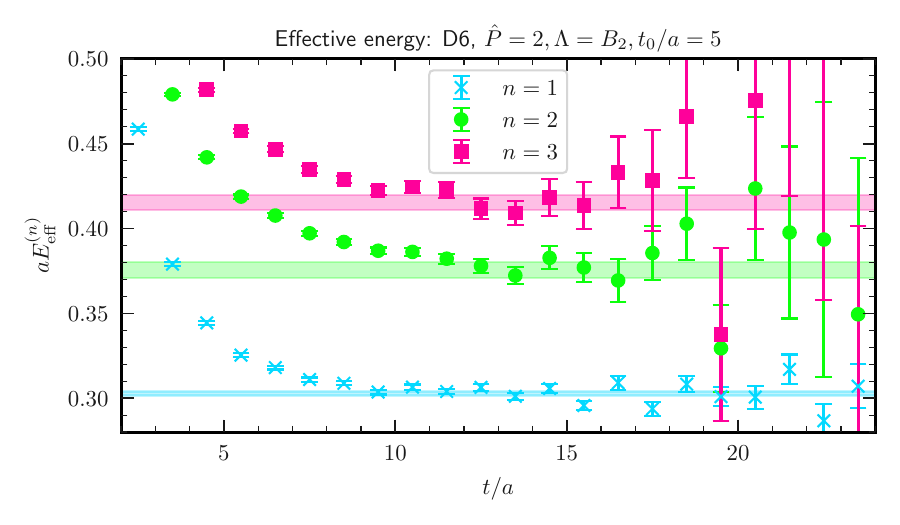}
\caption{Comparison of effective masses to the AIC extracted energy levels. The points with error bars show the lattice data. The light blue crosses, green dots, and pink squares show effective masses calculated from the principal correlators with $n=1,2,3$, respectively. The shaded region represents the $1\sigma$ region for each extracted energy level from the AIC. Plots for the other irreducible representations are given in the Supplemental Material. \label{fig:meff}}
\end{figure}

Each principal correlator decays exponentially in time, corresponding to the $n$-th lowest energy level of the correlation matrix. However, each correlator also includes excited-state contamination arising from energies outside the region covered by the interpolators. We can visualize this by plotting effective energies. To calculate $E_{\rm eff}$ we use the discretized formula:
\begin{align}
    aE_{\rm eff} (t+0.5a) = \ln \left[ \frac{\lambda(t)}{\lambda(t+a)} \right]  .
\end{align}
In figure \ref{fig:meff}, we plot an example of effective masses on the \texttt{D6} ensemble. The figure shows the three energy levels for $ \hat{P} = 2$, $\Lambda = B_2$, and $t_0/a = 5$. The curvature at low $t$ reflects excited state contamination at low $t$ and the noise increases at larger $t$. Other examples of effective masses are in the supplemental material.

\begin{figure}[htb!]
\centering
\includegraphics[width=\textwidth]{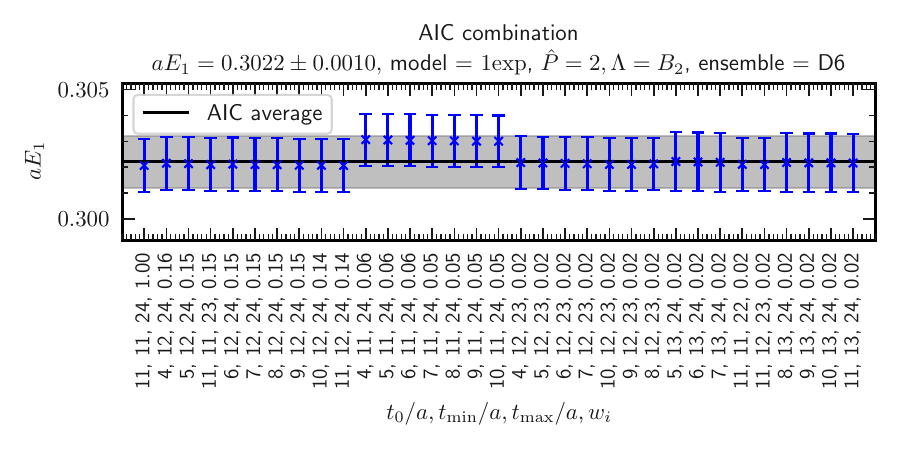}
\includegraphics[width=\textwidth]{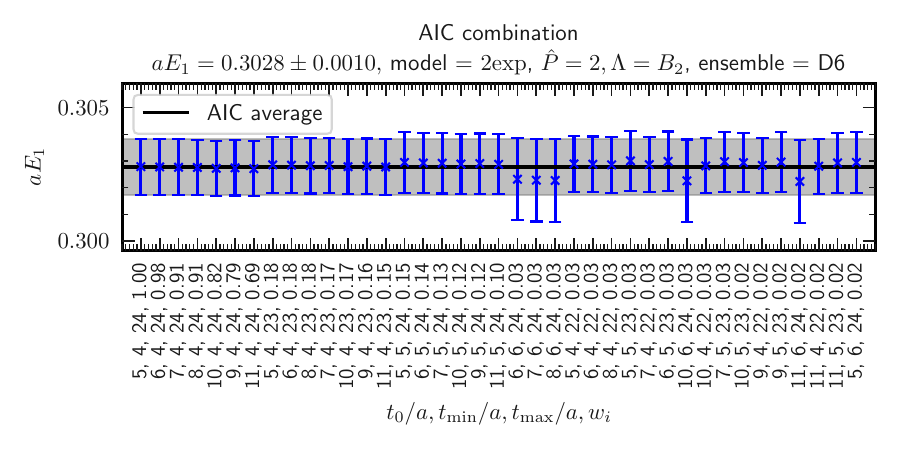}
\caption{Example of AIC combination for the lowest energy in the irreducible representation $B_2$ for $\hat{P}=2$. The blue crosses with error bars show fit results from individual models, and the black line with the shaded region shows the AIC combination of the models and the jackknife uncertainty. The horizontal axis includes reference time $t_0$, fit range $[t_{\rm min}, t_{\rm max}]$ and the AIC weight $w_i$ from equation \ref{eq:AICw}. For clarity, we rescaled the AIC weights in the labels so that the highest weight starts at $1.0$, and we show only models with weights above $0.01$. \label{fig:AIC}}
\end{figure}

We fit each principal correlator in two ways. First, with a single exponential:
\begin{align}
    \lambda_{n}(t,t_0) = A \exp{\left[-E_{n,1\rm exp} \left(t-t_0\right) \right]}, \label{eq:1exp} 
\end{align}
that holds for sufficiently large $t$, where contributions from higher energies are exponentially suppressed. The second ansatz contains two exponentials:
\begin{align}
    \lambda_{n}(t,t_0) = (1-B) \exp{\left[-E_{n,2\rm exp} \left(t-t_0\right) \right]} + B \exp{\left[-E_m \left(t-t_0\right) \right]}, \label{eq:2exp} 
\end{align}
which attempts to describe residual GEVP-excited-state contamination arising from outside the energy region covered by the interpolators. 

We perform multiple fits with each parameterization, varying $t_0$ and the fit range $[t_{\rm min},t_{\rm max}]$. The energy parameter $E_n$ is then averaged over $t_0$ and fit range, using the ``perfect model'' Akaike information criterion (AIC) \cite{Jay:2020jkz} for each parameterization separately. We calculate the AIC score for each fit,
\begin{align}
    s_{\rm AIC} = \frac{1}{N_{\rm jack}}\sum_{h=1}^{N_{\rm jack}} \left(\chi^2_h - 2\ \text{dof}\right),
\end{align}
where $\text{dof} = N_{\rm data} - N_{\rm parameters}$ and the average runs over the jackknife set. We discard the fits with $\chi^2/\text{dof}>1.5$ or $\chi^2/\text{dof}<0.4$, as well as the fits with an AIC score significantly higher than the minimum, $ s_{\text{AIC}} - s_{\rm AIC,min} > 15$. From the scores, we can calculate the weights and the weighted average:
\begin{align}
    w_{i} &= \frac{\exp\left[-0.5 s_{\text{AIC},i}\right]}{\sum_j \exp\left[-0.5 s_{\text{AIC},j}\right]},  \label{eq:AICw} \\
    E^{\hat{P},\Lambda}_{n,h,\text{AIC}} &= \sum_i w_{i} E^{\hat{P},\Lambda}_{n,h,i}. 
\end{align}
The indices $i,j$ run over all fitted models, and the index $h$ is the jackknife index. A representative example is shown in figure \ref{fig:AIC}; the rest can be found in the supplemental material. Figure \ref{fig:AIC} shows two AIC combinations for \texttt{1exp} and \texttt{2exp} parameterizations of the same energy level on the \texttt{D6} ensemble. Among the highest-weight fits, both parameterizations have similar $t_0$ and $t_{\rm max}$ distributions, while $t_{\rm min}$ is significantly lower for \texttt{2exp} at $t^{\rm 2exp}_{\rm min}/a \approx 5$ and $t^{\rm 1exp}_{\rm min}/a \approx 12$.

Throughout the analysis, we primarily use the two exponential parameterizations as the reference parameterization that dictates the central value and statistical error, both calculated from the jackknife resampled set of configurations:
\begin{align}
    \left<E_{x,\text{AIC}}\right> &= \frac{1}{N_{\rm jack}}\sum_{h=1}^{N_{\rm jack}} E_{x,h,\text{AIC}}, \\
    C_{xy,{\rm ens},{\rm stat}} &= \frac{N_{\rm jack}-1}{N_{\rm jack}} \sum_{h=1}^{N_{\rm jack}} \left(E_{x,h,\text{AIC}}- \left<E_{x,\text{AIC}}\right>\right)\left(E_{y,h,\text{AIC}}- \left<E_{y,\text{AIC}}\right>\right). 
\end{align}
The indices $x,y$ are triplet indices $(\hat{P}, \Lambda, n)$ and label the energy levels within the same ensemble. The single-exponential is used to estimate the systematic error:
\begin{align}
    \sigma_{x,\text{sys}}^2 = \max\left[\left|\sigma_{x,\text{1exp}}^2-\sigma_{x,\text{2exp}}^2\right|,\left|E_{x,\text{1exp}}-E_{x,\text{2exp}}\right|^2\right]. \label{eq:sysE}
\end{align}
The full covariance matrix $C_{ens}$ for each of the four ensembles is constructed as:
\begin{align}
    C_{xy,\rm ens} =  C_{xy,{\rm ens},{\rm stat}} + \delta_{x,y} \sigma_{x,\rm sys}^2. \label{eq:covEns}
\end{align}

\section{Energy spectrum results} \label{sec:Espec}
\begin{figure}[htbp]
\centering
\includegraphics[width=\textwidth]{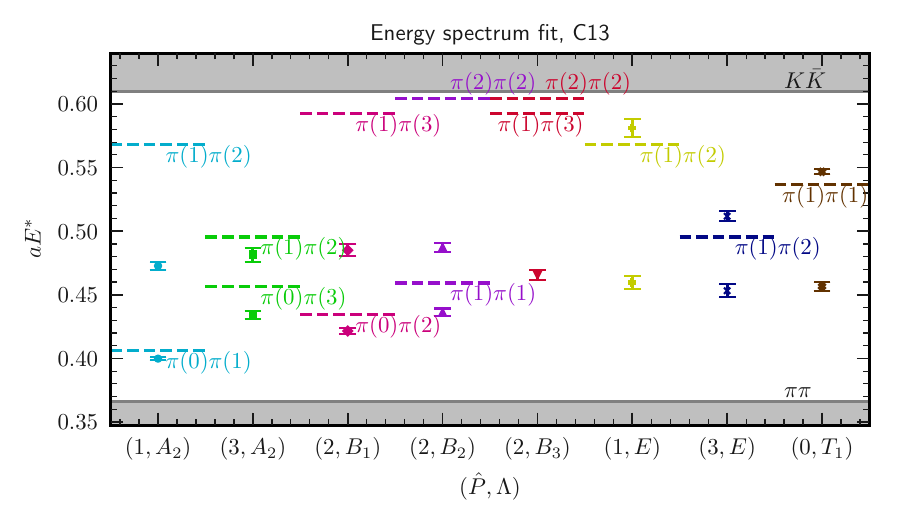}
\includegraphics[width=\textwidth]{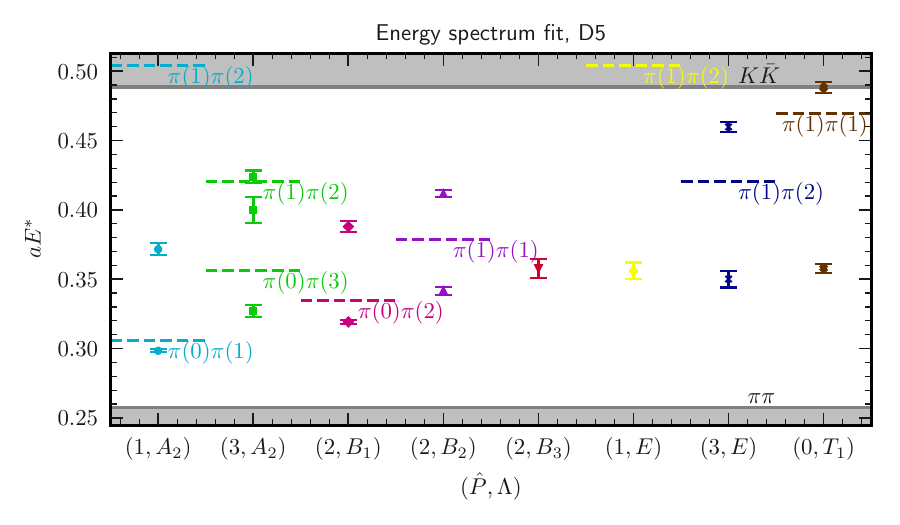}
\caption{Energy spectrum for \texttt{C13} (top) and \texttt{D5} (bottom) ensembles, between $\pi\pi$ and $KK$ thresholds. Dashed lines represent the noninteracting energies of two pions. \label{fig:spec12}}
\end{figure}

\begin{figure}[htbp]
\centering
\includegraphics[width=\textwidth]{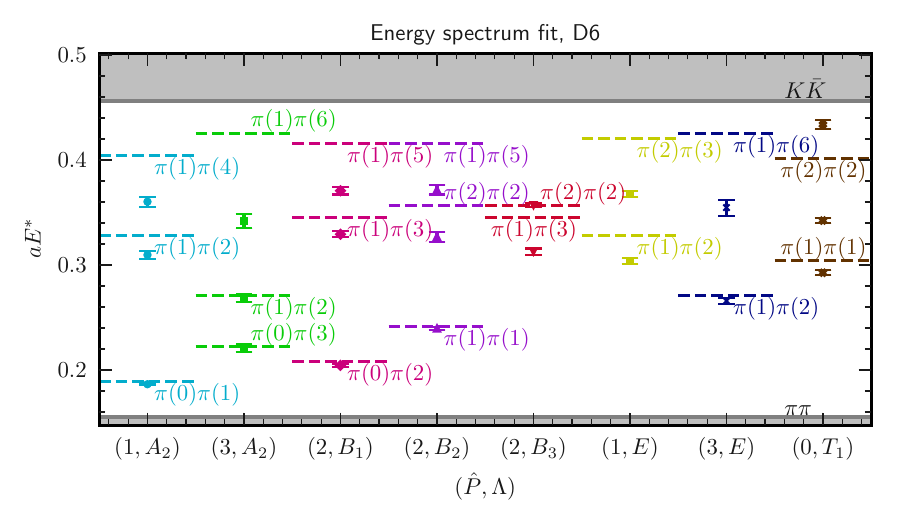}
\includegraphics[width=\textwidth]{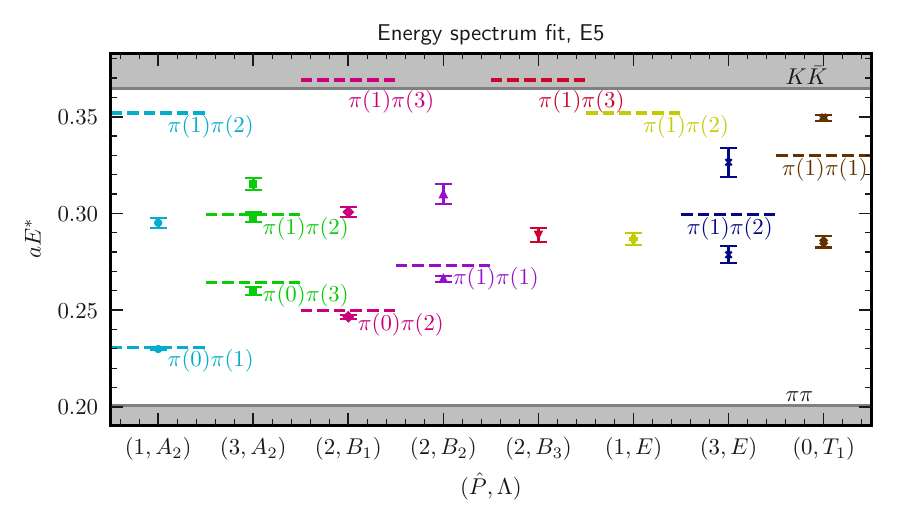}
\caption{ Same as figure \ref{fig:spec12}, for \texttt{D6} (top) and \texttt{E5} (bottom) ensembles.\label{fig:spec34}}
\end{figure}

The extracted energies are shown with error bars in figures \ref{fig:spec12} and \ref{fig:spec34}. The horizontal axis shows the irreducible representation and total momentum, $(\hat{P}, \Lambda)$, and the vertical axis shows the lattice energy in the center-of-momentum (CM) frame, $aE^* = \sqrt{(aE)^2 - (a|\textbf{P}|)^2}$. The noninteracting energies are shown with dashed lines, and the text beside the labels, $\pi(\hat{p}_1)\pi(\hat{p}_2)$, indicates the size of the momenta of the respective pions. Note that in the \texttt{D6} part of figure \ref{fig:spec34}, there are lines for noninteracting energy levels with pion momenta larger than $\hat{p} =|\mathbf{p}|^2 \left(\frac{2\pi}{L}\right)^{-2} > 3$. As we only include pions with momenta $\hat{p} \leq 3$, these energy levels put an upper boundary on the energies we can extract. 

In a finite volume, the presence of an elastic resonance like the $\rho$ manifests itself through the characteristic modification of the two-hadron energy spectrum. In the absence of interactions  the phase shift describing the two-pion interactions is zero, which gives rise to the noninteracting energy levels; however, when we have a resonance such as the $\rho$, where the phase shift rises from $0$ to $\pi$, the energies near the noninteracting state shift away from the noninteracting energy levels according to the Lüscher quantization condition \cite{Luscher:1990ux}. In the single-channel and single-partial-wave case, the quantization condition can be written schematically as
\begin{align}
    \label{eq:simpleQC}
    \cot{\delta(E)} = \cot{\phi(E,L)}
\end{align}
where $\cot{\phi}$ is a linear combination of Zeta functions \cite{Rummukainen:1995vs, Leskovec:2012gb}. When the phase shift $\delta$ exhibits a resonance, equation \eqref{eq:simpleQC} gives rise to an additional solution compared to the noninteracting spectrum. Depending on the nature of the resonance, the extra energy level can lie near the naively expected particle mass, or far away, and in certain cases it does not appear at all \cite{Lang:2012sv}. In the case of the $\rho$, there is an additional energy level present. 

Of the $66$ energy levels shown in figures \ref{fig:spec12} and \ref{fig:spec34}, we exclude $9$ in the global fit for the following reasons. First, we exclude $4$ energies: $\hat{P}=3, \Lambda=A_2, n=2$ from \texttt{D6}, $\hat{P}= 2, \Lambda=B_3, n=2$ from \texttt{D6}, $\hat{P}= 1, \Lambda=A_2, n=1$ from \texttt{E5}, and $\hat{P}= 3, \Lambda=A_2, n=2$ from \texttt{E5}, as they are too close to the noninteracting energy levels. Our quantization condition, described in the next section, has poles at noninteracting energy levels. The extracted energies are roots of the quantization condition, and fitting roots that are very close to the pole introduces two problems. First, the fit is very sensitive to slight perturbations of these energies, as the derivative is very large. Second, if the root is on top of the pole or on the wrong side, the root-finding algorithm cannot find it, and the fit fails.

The remaining $5$ excluded energies are $\hat{P}=3, \Lambda=E, n=2$, \texttt{D5}, $\hat{P}=0, \Lambda=T_1, n=2$, \texttt{D5}, $\hat{P}=2, \Lambda=B_1, n=3$, \texttt{D6}, $\hat{P}=0, \Lambda=T_1, n=3$, \texttt{D6}, and $\hat{P}=0, \Lambda=T_1, n=2$, \texttt{E5}. These levels include uncontrolled excited state contamination that we were unable to describe. We discuss these energies in more detail in the appendix \ref{ap:amps}. 

An \texttt{HDF5} file containing all energy values used in the global fit and their covariances can be found in the Supplemental Material.

%%%%%%%%%%%%%%%%%%%%%%%%%%%%%%%%%%%%%%%%%%%%%%%%%%%%%%%%%%%%%%%%%%%%%%%%%%%%%%%%%%%%%%%%%%%%%%%%%%%%%%%%%%%%%%%%%%%%%%%%%%%%%%%%%%%%%%%%%%%%%%%%%%%%%%%%%%%%%%%%%%%%%%%%%%%%%%%%%%%%%%%%%%%%%%%%%%%%%%%%%%%%%%%%%%%%%%%%
\section{Global-fit method}

We perform a global fit of the scattering amplitude to the energy spectra of all four ensembles. This fit includes parameterizing the quark-mass dependence of the K matrix, solving the finite-volume quantization conditions, and modeling discretization effects in the energy levels. To perform the global fit, we order all energies in a single vector
\begin{align}
    E_{\rm data} = \left\{\{E_n^{\hat{P}\Lambda}\}_{\texttt{C13}},\{E_n^{\hat{P}\Lambda}\}_{\texttt{D5}},\{E_n^{\hat{P}\Lambda}\}_{\texttt{D6}},\{E_n^{\hat{P}\Lambda}\}_{\texttt{E5}} \right\}
\end{align}
that contains energy spectra from all four ensembles. From the individual covariance matrices, using equation (\ref{eq:covEns}), we assemble a block-diagonal covariance matrix:
\begin{align}
    C = \text{diag}\{C_{\texttt{C13}},C_{\texttt{D5}},C_{\texttt{D6}},C_{\texttt{E5}}\},    
\end{align}
where we assume that the energy levels from different ensembles are uncorrelated. The global $\chi^2$ is then given by
\begin{align}
    \chi^2_\text{global fit} = \left(E_{\rm data} - E_{\rm model}\right)^T C^{-1} \left(E_{\rm data} - E_{\rm model}\right). \label{eq:chi2}
\end{align}

\subsection{Discretization effects} \label{sec:discretization}
Our quantization condition is derived in the framework of a finite-volume, continuum quantum field theory, so we model the discretization effects by splitting the $E_{\rm model}$ into two contributions in the CM frame:
\begin{align}
    E^*_{\rm model} = E^*_T - E^*_d. \label{eq:Eshift}
\end{align}
$E^*_T$ is the set of energies obtained by solving the quantization condition in the continuum limit, and $E^*_d$ is the correction that models the discretization effects by fitting the lattice spacing dependence of the energy spectrum. These discretization effects are expected to primarily arise from the discretization of fermion propagators, and therefore scale quadratically with the lattice spacing, given our use of the clover action. We include three terms:
\begin{itemize}
    \item an $(a|\mathbf{P}|)^2$ term that allows for discretization errors dependent on the total momentum,
    \item an $(ak)^2$ term that allows for discretization errors dependent on the scattering momentum
    \item a term $(a\Lambda_{\rm QCD})^2$, where $\Lambda_{\rm QCD} = 300$ MeV, to describe momentum-independent discretization effects.
\end{itemize}

With these three terms, we construct 12 different models: 6 additive and 6 multiplicative. First, the additive models: 
\begin{align}
    a E_{d,a1} &= c_C a (a\Lambda_{\rm QCD})^2, \label{eq:disc-a1} \\ 
    a E_{d,a2} &= c_P a (a|\mathbf{P}|)^2, \label{eq:disc-a2} \\ 
    a E_{d,a3} &= c_k a (ak)^2,\\
    a E_{d,a4} &= c_P a (a|\mathbf{P}|)^2 + c_k a (ak)^2,\\
    a E_{d,a5} &= c_P a (a|\mathbf{P}|)^2 + c_C a (a\Lambda_{\rm QCD})^2,\\
    a E_{d,a6} &= c_P a (a|\mathbf{P}|)^2 + c_k a (ak)^2  + c_C a (a\Lambda_{\rm QCD})^2.
\end{align}
The multiplicative models use the same structure but scale with energy:
\begin{align}
    a E_{d,m1} &= aE^*_T \left[c_C (a\Lambda_{\rm QCD})^2\right], \\
    a E_{d,m2} &= aE^*_T \left[c_P (a|\mathbf{P}|)^2\right], \\
    a E_{d,m3} &= aE^*_T \left[c_k (ak)^2\right], \\
    a E_{d,m4} &= aE^*_T \left[c_P (a|\mathbf{P}|)^2 + c_k (ak)^2\right], \\
    a E_{d,m5} &= aE^*_T \left[c_P (a|\mathbf{P}|)^2 + c_C (a\Lambda_{\rm QCD})^2\right], \\
    a E_{d,m6} &= aE^*_T \left[c_P (a|\mathbf{P}|)^2 + c_k (ak)^2 + c_C (a\Lambda_{\rm QCD})^2\right] \label{eq:disc-m6}.
\end{align}
Above, $c_C$, $c_P$, and $c_k$ are the fit parameters.

\subsection{Quantization condition} \label{ssec:QC}
The set of energies $\{E^*_T\}$ is obtained by solving the quantization condition in the continuum limit \cite{Luscher:1990ux, Guo:2012hv}: 
\begin{align}
    \det\left\{ c_{\Lambda,lm} \left[ T_{lm,l'm'}^{-1}(E_T^*) + F_{lm, l'm'} (E_T^*) \right] c^\dagger_{l'm',\Lambda} \right \}= 0, \label{eq:QC} 
\end{align}
where $E^* = \sqrt{s}$ is the CM energy. Here, $c_{\Lambda,lm}$ are projection matrices that project from spherical harmonics to irreducible representations $\Lambda$. We compute them in the same way we compute projections for the correlation functions, as described in section \ref{sec:2pt}. The fit includes parameterizing the infinite volume scattering amplitude $T$ and using a root-finding algorithm to find the roots of equation (\ref{eq:QC}). The algorithm returns a set of energies $\left\{E^*_T\right\}$ for a given parameterization of $T(E_T^*)$. 

The function F is a geometric function as described in \cite{arXiv:1706.06223}:
\begin{align}
    F_{lm, l'm'} = i \rho \left[ \delta_{l,l'}\delta_{m,m'} + i \frac{E^*}{\pi^{3/2} E} \sum_{\bar{l},\bar{m}} b\left(l,m; \bar{l},\bar{m}|l'm'\right) \left(\frac{kL}{2\pi}\right)^{-\bar{l}-1} \mathcal{Z}_{\bar{l},\bar{m}}^d\left(1; \frac{k^2L^2}{4\pi^2}\right) \right],
\end{align}
where $\rho = \frac{k}{16\pi \sqrt{s}}$, $b$ is a Clebsch-Gordan term, and $\mathcal{Z}$ is a Lüscher Zeta function. The latter are described in detail in the last section of ref.~\cite{arXiv:1706.06223}. 

The amplitude $T$ satisfies the unitarity condition $T-T^\dagger = 2i T \rho \ T^\dagger$. One way to parameterize T in a unitary manner is with a real and symmetric matrix $K$ \cite{martin1970elementary}:
\begin{align}
    T^{-1} &= K^{-1} - i\rho.  \label{eq:T-K}
\end{align}
We divide out the centrifugal barrier and get the ``reduced $K$ matrix'', denoted as $\tilde{K}$:
\begin{align}
    K_{lm, l'm'} &= \tilde{K}_{l} \delta_{ll'} \delta_{mm'} k^{2l}. 
\end{align}
Our parameterization of $\tilde{K}_1$ includes a resonant pole and a polynomial expansion in powers of $s$:
\begin{align}
    \tilde{K}_1 = \frac{g^2}{m^2_\rho -s} + \sum_{n=0}^{n_{\rm max}} c_n s^n.
\end{align}
The resonant pole $\frac{g^2}{m^2_\rho -s}$ reproduces the Breit-Wigner form $\rho T_{11} = \frac{\sqrt{s}\Gamma(s)}{m_\rho^2 -s-i\sqrt{s}\Gamma(s)}$  from \cite{Alexandrou:2017mpi}, where $\Gamma$ is a function of $s$. The polynomial terms in $\tilde{K}$ are used to probe sensitivity to smooth contributions from higher-lying resonances. These terms, together with the inclusion of the $l=3$ partial wave, are discussed in appendix~\ref{ap:amps}. For all the fits in section \ref{sec:res}, we used only the term with the pole:
\begin{align}
    \tilde{K} = \frac{g^2(m_\pi,m_{\eta_s})}{m_\rho^2(m_\pi,m_{\eta_s}) - s}, \label{eq:Kmat}
\end{align}
with $m_\rho$ and $g^2$ as functions of $m_\pi$ and $m_{\eta_s}$, which are used as proxies for the light and strange quark masses. These functions depend on fit parameters, as discussed in section \ref{sec:mass-dep}. 

A convenient way to present the elastic scattering amplitude is in terms of the phase shift. The scattering amplitude is related to the phase shift via
\begin{align}
    \delta_l = \arctan \left[ \frac{1}{T_{ll}^{-1} \rho^{-1} + i } \right]. \label{eq:psT}
\end{align}
In our main analysis, we only work in the $l=1$ subspace, and we suppress the $l$ index. We can also drop the determinant in equation \ref{eq:QC} and rewrite for $F$:
\begin{align}
    -F(E^*_T) &= T^{-1}(E^*_T), \\
    \delta(E^*_T) &= \arctan \left[ \frac{1}{-F(E^*_T) \rho^{-1} + i } \right]. \label{eq:psF}
\end{align}
By comparing the phase shift from the fitted scattering amplitude $T$ to the phase shift calculated from the data points of the energy spectra, shifted to the continuum limit $E_T^* = E^*_{\rm data} + E_d^*$, we will visualize the fits in section \ref{sec:res}.

\subsection{Mass dependence} \label{sec:mass-dep}

\begin{figure}[htbp]
\centering
\begin{subfigure}{0.3\textwidth}
    \centering
    \includegraphics[width=\textwidth]{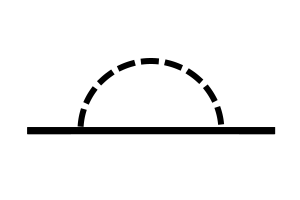}
    \caption{}
\end{subfigure}
\begin{subfigure}{0.3\textwidth}
    \centering
    \includegraphics[width=\textwidth]{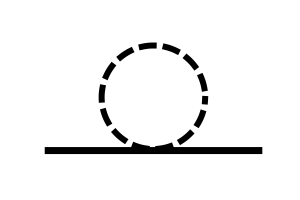}
    \caption{}
\end{subfigure}
\begin{subfigure}{0.3\textwidth}
    \centering
    \includegraphics[width=\textwidth]{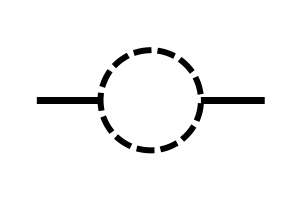}
    \caption{}
\end{subfigure}
\caption{ Diagrams contributing to the $\rho$ mass in the approach of ref.~\cite{Bruns:2004tj}: a) sunrise diagram on the left, b) tadpole diagram in the middle, and c) self-energy diagram on the right.  The full line represents the $\rho$ meson, and the dashed line represents the $\pi$. \label{fig:diagrams}}
\end{figure}

The $K$-matrix parameters from equation (\ref{eq:Kmat}) depend on the quark masses of the ensembles. We parameterize this dependence with $m_\pi$ and $m_{\eta_s}$ mass dependence, where the $\eta_s$ is a fictitious flavor-nonsinglet meson consisting of two quarks with masses equal to the strange-quark mass \cite{HPQCD:2011qwj}. We parameterize the $\eta_s$-mass dependence with the quadratic term
\begin{align}
    r_{\eta_s}^2 = \frac{\left(am_{\eta_s}\right)^2 - \left(am_{\eta_s,\rm phys}\right)^2 }{\left(am_{\eta_s,\rm phys}\right)^2},
\end{align} 
where the physical $\eta_s$ mass is $m_{\eta_s,\rm phys} = 688.5$ MeV \cite{HPQCD:2011qwj}. 
For the $m_\pi$ dependence, we consider the following possible terms:
\begin{align}
    r_\pi^2 &= \frac{\left(am_\pi\right)^2 - \left(am_{\pi,\rm phys}\right)^2 }{\left(am_{\pi,\rm phys}\right)^2}, \label{eq:mpi2term}\\
    r_\pi^3 &= \frac{\left(am_\pi\right)^3 - \left(am_{\pi,\rm phys}\right)^3 }{\left(am_{\pi,\rm phys}\right)^3}, \\
    r_\pi^{2\log} &= \frac{\left(am_\pi\right)^2}{\left(am_{\pi,\rm phys}\right)^2} \ln{\frac{\left(am_\pi\right)^2}{\left(a\mu\right)^2}} - \ln{\frac{\left(am_{\pi,\rm phys}\right)^2}{\left(a\mu\right)^2}}, \\
    r_\pi^{4\log} &= \frac{\left(am_\pi\right)^4}{\left(am_{\pi,\rm phys}\right)^4} \ln{\frac{\left(am_\pi\right)^2}{\left(a\mu\right)^2}} - \ln{\frac{\left(am_{\pi,\rm phys}\right)^2}{\left(a\mu\right)^2}}, \label{eq:mpi2+4term}
\end{align}
where we set the constant $\mu = 1$ GeV. These terms are motivated by the prediction of Bruns-Meissner \cite{Bruns:2004tj}:
\begin{align}
    m_\rho = m_{\rho,0} + c_1 m^2_\pi + c_2 m^3_\pi + c_3 m^4_\pi \ln{\frac{m^2_\pi}{m^2_\rho}} + \mathcal{O}\left(m^4_\pi\right),
\end{align}
derived within a chiral EFT.  The first term, $m_{\rho,0}$, is the $\rho$ mass in the chiral limit, at zero quark mass. The quadratic term $c_1 m^2_\pi$ comes from operators containing the quark mass matrix. The $c_2 m^3_\pi$ term is produced by the sunrise diagram drawn in figure \ref{fig:diagrams}.a. This diagram gives two contributions $k_1 m^3_\pi$ and  $k_2 m^4_\pi \ln{\frac{m^2_\pi}{m^2_\rho}}$. This is the only diagram with the $m^3_\pi$ contribution. The term $c_3 m^4_\pi \ln{\frac{m^2_\pi}{m^2_\rho}}$ has two other contributions from one-loop diagrams. Alongside the sunrise diagram, we get a contribution from the tadpole diagram (\ref{fig:diagrams}.b) and a contribution from the self-energy diagram (\ref{fig:diagrams}.c) with a pure Goldstone-boson loop.

The $r_\pi^{2\log}$ term lacks a clear physical motivation. We decided to keep it anyway, as it can be used to improve the estimation of systematic uncertainties. For the physical pion mass, we adopted the charged-pion value $m_{\pi,\rm phys} = 139.57039 (17)$ MeV \cite{ParticleDataGroup:2026mpi}, despite our theory lacking isospin breaking and electromagnetic interactions. This value is most appropriate for applications that involve $\rho^0$, which decays to $\pi^+\pi^-$ only.

In the scattering amplitude, there are two parameters: $m_\rho$ and $g^2$. The derivation by Bruns-Meissner was done only for the dependence of $m_\rho$ on the pion mass. To model the dependence of $g^2$ on the quark masses, we decided to use the same functional forms as used for $m_\rho$, but with independent parameters. From previous studies \cite{Bali:2015gji}, we expect $g^2$ to have very weak quark-mass dependence. 

With four ensembles in the global fit we can include at most four parameters each in our mass-dependence parameterizations of $g^2$ and $m_\rho$. One parameter is the physical value. We then add up to two $\pi$-mass-dependence terms and the quadratic $\eta_s$-mass dependence to each parameter of $\tilde{K}$. 
In the global fit we consider 7 combinations:
\begin{align}
    p_1 &= p_{\rm phys} + c_{p,2} r^2_\pi + c_{p,\eta_s} r_{\eta_s}^2, \label{eq:mass-p1}\\
    p_2 &= p_{\rm phys} + c_{p,2\rm log} r^{2\rm log}_\pi + c_{p,\eta_s} r_{\eta_s}^2,\\
    p_3 &= p_{\rm phys} + c_{p,3} r^3_\pi + c_{p,\eta_s} r_{\eta_s}^2,\\
    p_4 &= p_{\rm phys} + c_{p,4\rm log} r^{4\rm log}_\pi + c_{p,\eta_s} r_{\eta_s}^2,\\
    p_5 &= p_{\rm phys} + c_{p,2} r^2_\pi + c_{p,2\rm log} r^{2\rm log}_\pi+ c_{p,\eta_s} r_{\eta_s}^2,\label{eq:mass-p5}\\
    p_6 &= p_{\rm phys} + c_{p,2} r^2_\pi + c_{p,3} r^3_\pi + c_{p,\eta_s} r_{\eta_s}^2,\\
    p_7 &= p_{\rm phys} + c_{p,2} r^2_\pi + c_{p,4\rm log} r^{4\rm log}_\pi + c_{p,\eta_s} r_{\eta_s}^2, \label{eq:mass-p7}
\end{align}
where $p$ stands for either $m_\rho$ or $g^2$. Each fitted mass-dependence model always uses the same parameter combination $p_x$ for both $m_\rho$ and $g^2$, with different coefficients $c_{p,x}$. 

\section{Global-fit results} \label{sec:res}

Here we discuss the results of the simultaneous fits to the energy spectra for every combination of discretization effect models described in section \ref{sec:discretization} and mass dependence models described in section \ref{sec:mass-dep}. All parameter values and covariance matrices from the fitted models are provided in the supplemental material. First, we discuss the phase shifts for each ensemble for a specific set of models, and then we compare them with the experimental results. We then examine the pion mass and chiral-continuum extrapolation of the $\rho$ pole, and assess its sensitivity to the mass parameterization; finally, we combine the results from all models and average them. 

\subsection{Phase shifts}

\begin{figure}[htbp]
\centering
\includegraphics[width=\textwidth]{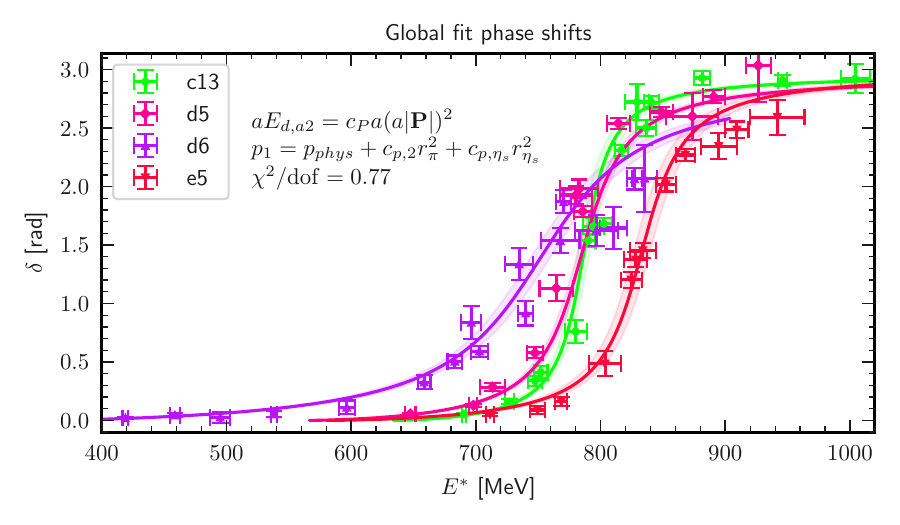}
\caption{Continuum phase shifts for the unphysical $\pi$ and $\eta_s$ masses on each ensemble, from the global fit with the discretization model $aE_{d,a2}$ and mass dependence model $p_1$. The phase-shift data points shown here are obtained from the energy levels after subtracting the modeled discretization effects. \label{fig:pshift}}
\end{figure}

Figure \ref{fig:pshift} shows an example of our global fits, using the $aE_{d,a2}$ parameterization [equation (\ref{eq:disc-a2})] for the discretization effects and the $p_1$ parameterization [equation (\ref{eq:mass-p1})] for the light- and strange-quark mass dependence. We show the phase shift curves evaluated for the $m_\pi$ and $m_{\eta_s}$ values of each ensemble, along with the phase shift points corresponding to each energy level, corrected for the discretization effects [these phase-shift points are given by equation (\ref{eq:psF})].
The phase shifts for the remaining model combinations are qualitatively similar and are presented in the supplemental material. 

When comparing ensembles \texttt{D6}, \texttt{D5}, and \texttt{C13}, we observe an increase in the phase-shift turnover point (where the phase reaches $\pi/2$) as the pion mass increases. At the same time, the slope becomes steeper, indicating a narrowing of the $\rho$ decay width, as the phase space becomes smaller. Note that the ensemble \texttt{E5} has a similar pion mass to \texttt{D5}, yet the turnover point appears shifted towards higher energies. Our fit model attributes this effect to the strange-quark-mass dependence. Comparing the $\eta_s$ masses on all four ensembles, see Table \ref{tab:ens}, reveals that \texttt{C13}, \texttt{D5}, and \texttt{D6} all have a value close to the physical $\eta_s$ \cite{Davies:2009tsa}, while on \texttt{E5} it is much higher.

\begin{figure}[htbp]
\centering
\includegraphics[width=\textwidth]{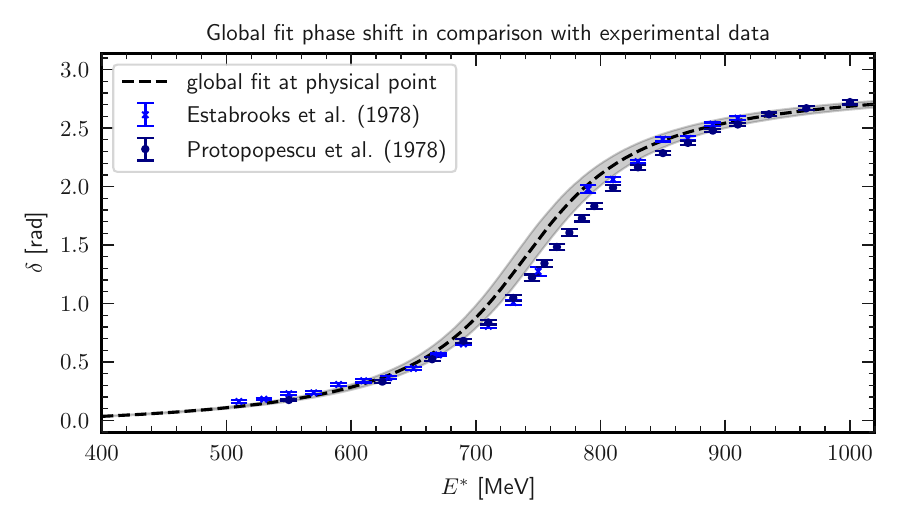}
\caption{Phase shift for the global fit with the discretization model $aE_{d,a2}$ and mass dependence model $p_1$ in the continuum and in the physical-quark-mass limit, compared to the experimental phase shifts from the papers \cite{Estabrooks:1974vu} and \cite{Protopopescu:1973sh}. \label{fig:pshift2}}
\end{figure}

In figure \ref{fig:pshift2} we compare the extrapolation to the physical quark masses, both light and strange, to the experimentally measured phase shifts of \cite{Estabrooks:1974vu, Protopopescu:1973sh}; the lower and higher energy regions match well, while the mid region shows minor tension with Protopopescu et al. \cite{Protopopescu:1973sh}, while some points from Estabrooks et al. \cite{Estabrooks:1974vu} match perfectly. 

\subsection{$\rho$ pole trajectories}

\begin{figure}[htbp]
\centering
\includegraphics[width=\textwidth]{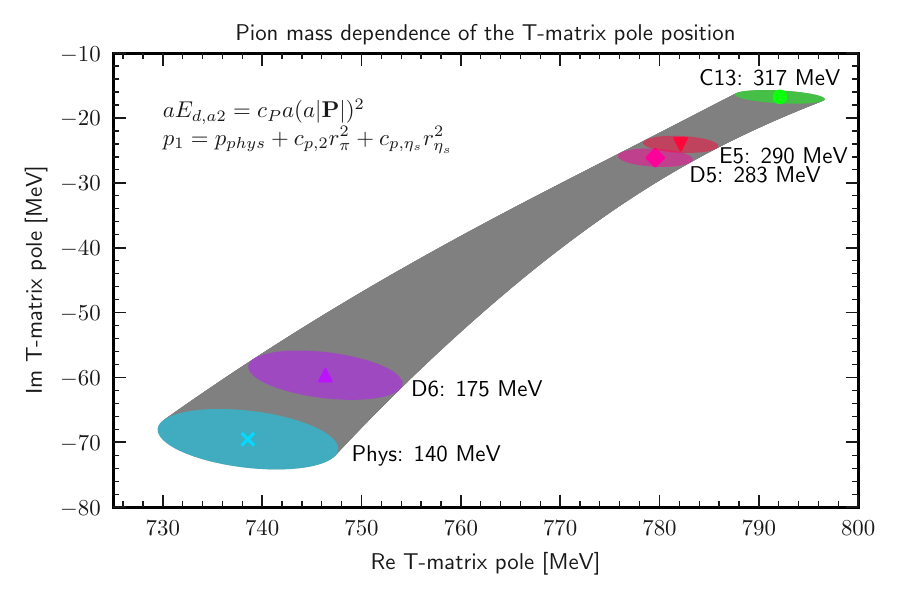}
\caption{Dependence of the complex pole position on the pion mass, at the physical strange-quark mass. The fit uses the discretization model $aE_{d,a2}$ and the mass-dependence model $p_1$. \label{fig:pole-pimass}}
\end{figure}

The shape of the $\pi\pi$ $I=1$, $P$-wave phase shift is dictated by the location of the complex pole attributed to the $\rho$ resonance in the scattering amplitude $T$. We determine the pole by performing a numerical search for zeros of the model $T^{-1}$; we pick a starting point close to the physical $\rho$ pole and search for the nearest complex zero. An example of the pion-mass dependence of the complex pole is shown in figure \ref{fig:pole-pimass}, where we plot the continuum pole position at different pion masses corresponding to the pion masses of the ensembles as colored (\texttt{C13} - green, \texttt{E5} - red, \texttt{D5} - magenta and \texttt{D6} - purple) shaded regions; the physical-pion-mass extrapolation is shown in the teal color. The gray shaded region indicates the pole trajectory as the pion mass is changed.

\begin{figure}[htbp]
\centering
\includegraphics[width=\textwidth]{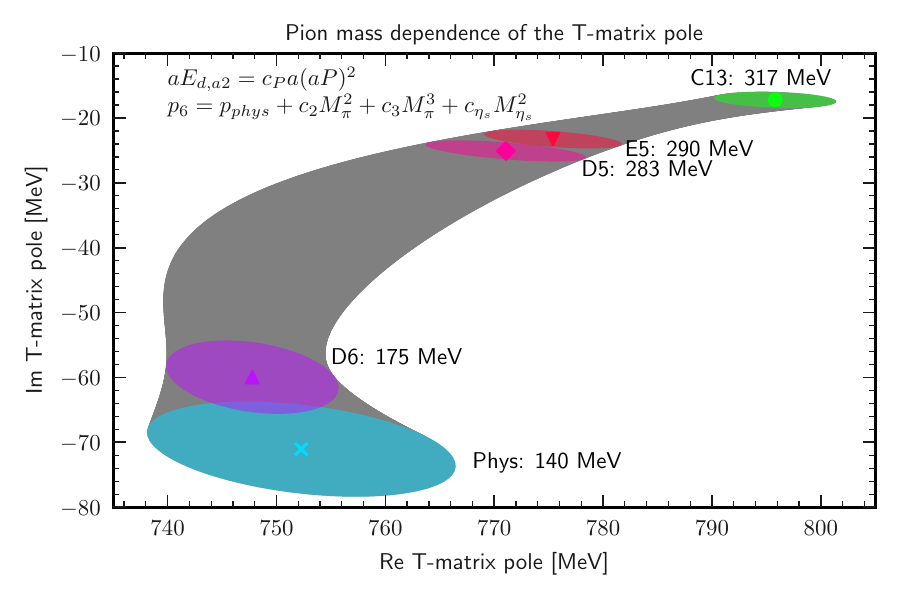}
\caption{Dependence of the complex pole on the pion mass, at the physical strange-quark mass. This fit uses the discretization model $aE_{d,a2}$ and the mass-dependence model $p_6$. \label{fig:pole-pimass2}}
\end{figure}

In certain cases, we observe a boomerang shape (see figure \ref{fig:pole-pimass2}), which occurs in four-parameter mass-dependence models. With four parameters fitted on four ensembles, we have no degrees of freedom in the mass dependence and thus much larger error bars on fitted parameters than in three-parameter models. The two $m_\pi$-dependent terms [equation (\ref{eq:mass-p5})-(\ref{eq:mass-p7})], can counteract at certain $m_\pi$ values, which allows for the real part of the pole to increase while $m_\pi$ decreases. We include these models in our analysis because they are equivalent to three-parameter models within $1\sigma$; however, to further assess their validity, we would need energy spectra from additional ensembles.
Additional pole trajectories are shown in the supplemental material.

\subsection{The $\rho$ pole at the physical point}

As the $\rho$ pole at physical quark masses is a common feature of all the quark-mass-dependence and lattice-spacing-dependence models, we can obtain a model-averaged result for the physical $\rho$ pole mass and decay width using the AIC \cite{Jay:2020jkz}. We show the model variation of the real and imaginary parts of the $\rho$ pole position in Table \ref{tab:pole-AIC}.
For each model, the pole uncertainty is obtained by propagating the joint fitted-parameter covariance determined from the Hessian.

The energy spectra of the four ensembles $\{E^*_{\rm data}\}$ are uncorrelated. In a fit without discretization effects, the models with four-parameter mass dependence $p_{5,6,7}$ would have the same $\chi^2/\text{dof}$. In Table \ref{tab:pole-AIC} we see that models with four-parameter mass dependence and the same discretization effects have similar $\chi^2/\text{dof}$, but not exactly the same. This shows that discretization-effect parameterizations introduce small correlations between ensembles in the continuum energy spectra $\{E^*_{\rm data} + E^*_d\}$.

\begin{table}[]
    \centering
    \setlength{\tabcolsep}{3pt}
    \renewcommand{\arraystretch}{0.85}

    \begin{tabular}{cc|ccc|@{\quad}|cc|ccc}
        $E_d^*$ & $p(m)$ &
        $\mathrm{Re}\sqrt{s}_\rho$ &
        $\mathrm{Im}\sqrt{s}_\rho$ &
        $\chi^2/\mathrm{dof}$
        &
        $E_d^*$ & $p(m)$ &
        $\mathrm{Re}\sqrt{s}_\rho$ &
        $\mathrm{Im}\sqrt{s}_\rho$ &
        $\chi^2/\mathrm{dof}$ \\
        
        \hline
        d,a1 & $p_1$ & $740 (10)$ & $-68 (7)$ & $0.80$ & 
        d,m1 & $p_1$ & $740 (9)$ & $-68 (4)$ & $0.80$ \\ 
        d,a2 & $p_1$ & $739 (9)$ & $-69 (4)$ & $0.77$ & 
        d,m2 & $p_1$ & $740 (9)$ & $-65 (7)$ & $0.79$ \\ 
        d,a3 & $p_1$ & $738 (10)$ & $-70 (7)$ & $0.80$ & 
        d,m3 & $p_1$ & $740 (10)$ & $-67 (4)$ & $0.80$ \\ 
        d,a4 & $p_1$ & $741 (10)$ & $-68 (6)$ & $0.79$ & 
        d,m4 & $p_1$ & $740 (10)$ & $-68 (6)$ & $0.79$ \\ 
        d,a5 & $p_1$ & $742 (9)$ & $-67 (5)$ & $0.77$ & 
        d,m5 & $p_1$ & $742 (9)$ & $-69 (7)$ & $0.78$ \\ 
        d,a6 & $p_1$ & $739 (10)$ & $-68 (6)$ & $0.77$ & 
        d,m6 & $p_1$ & $741 (9)$ & $-64 (5)$ & $0.80$ \\ 
        d,a1 & $p_2$ & $737 (10)$ & $-71 (5)$ & $0.82$ & 
        d,m1 & $p_2$ & $737 (10)$ & $-66 (6)$ & $0.83$ \\ 
        d,a2 & $p_2$ & $736 (10)$ & $-70 (6)$ & $0.78$ & 
        d,m2 & $p_2$ & $737 (10)$ & $-69 (7)$ & $0.79$ \\ 
        d,a3 & $p_2$ & $735 (11)$ & $-67 (7)$ & $0.82$ & 
        d,m3 & $p_2$ & $738 (10)$ & $-68 (7)$ & $0.82$ \\ 
        d,a4 & $p_2$ & $740 (11)$ & $-70 (5)$ & $0.82$ & 
        d,m4 & $p_2$ & $738 (10)$ & $-69 (8)$ & $0.81$ \\ 
        d,a5 & $p_2$ & $739 (11)$ & $-67 (7)$ & $0.78$ & 
        d,m5 & $p_2$ & $738 (10)$ & $-74 (8)$ & $0.79$ \\ 
        d,a6 & $p_2$ & $739 (12)$ & $-70 (5)$ & $0.79$ & 
        d,m6 & $p_2$ & $738 (11)$ & $-74 (5)$ & $0.81$ \\ 
        d,a1 & $p_3$ & $742 (9)$ & $-67 (7)$ & $0.79$ & 
        d,m1 & $p_3$ & $742 (9)$ & $-68 (8)$ & $0.79$ \\ 
        d,a2 & $p_3$ & $741 (8)$ & $-68 (5)$ & $0.76$ & 
        d,m2 & $p_3$ & $741 (9)$ & $-70 (6)$ & $0.77$ \\ 
        d,a3 & $p_3$ & $741 (9)$ & $-64 (6)$ & $0.80$ & 
        d,m3 & $p_3$ & $741 (9)$ & $-66 (6)$ & $0.79$ \\ 
        d,a4 & $p_3$ & $742 (9)$ & $-68 (7)$ & $0.77$ & 
        d,m4 & $p_3$ & $742 (9)$ & $-66 (6)$ & $0.78$ \\ 
        d,a5 & $p_3$ & $743 (9)$ & $-65 (6)$ & $0.76$ & 
        d,m5 & $p_3$ & $743 (9)$ & $-68 (6)$ & $0.77$ \\ 
        d,a6 & $p_3$ & $743 (9)$ & $-68 (6)$ & $0.77$ & 
        d,m6 & $p_3$ & $741 (9)$ & $-67 (7)$ & $0.77$ \\ 
        d,a1 & $p_4$ & $743 (9)$ & $-67 (8)$ & $0.79$ & 
        d,m1 & $p_4$ & $742 (9)$ & $-66 (8)$ & $0.79$ \\ 
        d,a2 & $p_4$ & $741 (8)$ & $-68 (7)$ & $0.76$ & 
        d,m2 & $p_4$ & $741 (8)$ & $-67 (7)$ & $0.76$ \\ 
        d,a3 & $p_4$ & $741 (9)$ & $-67 (8)$ & $0.79$ & 
        d,m3 & $p_4$ & $742 (9)$ & $-67 (7)$ & $0.79$ \\ 
        d,a4 & $p_4$ & $741 (9)$ & $-68 (6)$ & $0.77$ & 
        d,m4 & $p_4$ & $743 (9)$ & $-68 (7)$ & $0.78$ \\ 
        d,a5 & $p_4$ & $744 (10)$ & $-67 (8)$ & $0.75$ & 
        d,m5 & $p_4$ & $743 (9)$ & $-67 (6)$ & $0.76$ \\ 
        d,a6 & $p_4$ & $741 (9)$ & $-68 (8)$ & $0.76$ & 
        d,m6 & $p_4$ & $743 (9)$ & $-68 (8)$ & $0.78$ \\ 
        d,a1 & $p_5$ & $763 (17)$ & $-71 (8)$ & $0.78$ & 
        d,m1 & $p_5$ & $741 (16)$ & $-68 (6)$ & $0.83$ \\ 
        d,a2 & $p_5$ & $747 (16)$ & $-72 (9)$ & $0.76$ & 
        d,m2 & $p_5$ & $755 (16)$ & $-71 (7)$ & $0.75$ \\ 
        d,a3 & $p_5$ & $762 (18)$ & $-83 (9)$ & $0.78$ & 
        d,m3 & $p_5$ & $741 (17)$ & $-68 (6)$ & $0.83$ \\ 
        d,a4 & $p_5$ & $758 (17)$ & $-78 (7)$ & $0.75$ & 
        d,m4 & $p_5$ & $761 (17)$ & $-75 (6)$ & $0.76$ \\ 
        d,a5 & $p_5$ & $759 (17)$ & $-73 (6)$ & $0.74$ & 
        d,m5 & $p_5$ & $758 (16)$ & $-75 (10)$ & $0.74$ \\ 
        d,a6 & $p_5$ & $758 (17)$ & $-75 (9)$ & $0.73$ & 
        d,m6 & $p_5$ & $756 (17)$ & $-73 (7)$ & $0.75$ \\ 
        d,a1 & $p_6$ & $759 (14)$ & $-68 (8)$ & $0.79$ & 
        d,m1 & $p_6$ & $744 (14)$ & $-69 (7)$ & $0.82$ \\ 
        d,a2 & $p_6$ & $753 (14)$ & $-71 (7)$ & $0.75$ & 
        d,m2 & $p_6$ & $739 (14)$ & $-67 (8)$ & $0.81$ \\ 
        d,a3 & $p_6$ & $753 (16)$ & $-69 (8)$ & $0.78$ & 
        d,m3 & $p_6$ & $759 (16)$ & $-69 (6)$ & $0.78$ \\ 
        d,a4 & $p_6$ & $740 (15)$ & $-68 (7)$ & $0.82$ & 
        d,m4 & $p_6$ & $740 (14)$ & $-67 (6)$ & $0.82$ \\ 
        d,a5 & $p_6$ & $742 (15)$ & $-66 (6)$ & $0.80$ & 
        d,m5 & $p_6$ & $741 (15)$ & $-67 (8)$ & $0.81$ \\ 
        d,a6 & $p_6$ & $756 (15)$ & $-75 (7)$ & $0.74$ & 
        d,m6 & $p_6$ & $755 (15)$ & $-71 (8)$ & $0.75$ \\ 
        d,a1 & $p_7$ & $759 (17)$ & $-71 (11)$ & $0.77$ & 
        d,m1 & $p_7$ & $749 (15)$ & $-69 (8)$ & $0.80$ \\ 
        d,a2 & $p_7$ & $754 (14)$ & $-75 (7)$ & $0.74$ & 
        d,m2 & $p_7$ & $756 (14)$ & $-74 (7)$ & $0.74$ \\ 
        d,a3 & $p_7$ & $757 (16)$ & $-73 (8)$ & $0.76$ & 
        d,m3 & $p_7$ & $741 (15)$ & $-65 (7)$ & $0.84$ \\ 
        d,a4 & $p_7$ & $758 (16)$ & $-74 (7)$ & $0.76$ & 
        d,m4 & $p_7$ & $741 (15)$ & $-65 (7)$ & $0.83$ \\ 
        d,a5 & $p_7$ & $761 (15)$ & $-73 (9)$ & $0.73$ & 
        d,m5 & $p_7$ & $742 (15)$ & $-65 (7)$ & $0.82$ \\ 
        d,a6 & $p_7$ & $759 (16)$ & $-72 (8)$ & $0.74$ & 
        d,m6 & $p_7$ & $748 (15)$ & $-67 (7)$ & $0.77$ \\ 

    \end{tabular}
    \caption{ Locations of the $T$-matrix pole and values of $\chi^2/{\rm dof}$ for every combination of discretization-effects and mass-dependence model. The first column shows the discretization model index [equations (\ref{eq:disc-a1}-\ref{eq:disc-m6})]. The second column specifies the mass-dependence parameterization [equations (\ref{eq:mass-p1}-\ref{eq:mass-p7})]. The pole positions are given in MeV units.}
    \label{tab:pole-AIC}
\end{table}

The pole at the physical pion mass $m_{\pi,\rm phys} = 139.57$ MeV is
\begin{align}
    \sqrt{s}_\rho =  744 (11) (7) - i 69 (7) (3)\ \text{MeV},
\end{align}
where the first uncertainties are the AIC averages of the model uncertainties,
\begin{align}
    \sigma^2_{\rm first} = \sum_{i=1}^{N_{\rm models}} w_i \sigma_i^2,
\end{align}
and the second uncertainties describe the dispersion of the models:
\begin{align}
    \sigma^2_{\rm second} = \sum_{i=1}^{N_{\rm models}} w_i (\text{CV})_i^2 - \left(\sum_{i=1}^{N_{\rm models}} w_i (\text{CV})_{i}\right)^2.
\end{align}
Here $\text{CV}_{i}$ denotes the central value for either real or imaginary part of the pole obtained from model $i$, and the model uncertainties $\sigma_i$ are calculated from the Hessian matrices of the individual model fits. 

\section{Conclusion}

We have determined the $\rho$ resonance pole position at the physical point in $I=1$, $P$-wave $\pi\pi$ scattering from a single global fit to the finite-volume spectra of four $N_f=2+1$ clover-Wilson ensembles with lattice spacings between $0.068$ and $0.114$~fm and pion masses between $175$ and $317$~MeV.
A new feature of the analysis is the separation between finite-volume energies arising from the continuum scattering amplitude and corrections from discretization effects. 
The L\"uscher quantization condition is solved at zero lattice spacing for the quark masses and volume of each ensemble, and the discretization corrections are added to these continuum finite-volume energies. The continuum scattering amplitude is described by a single $K$ matrix whose parameters depend explicitly on $m_\pi$ and $m_{\eta_s}$.

We repeated the fit for $84$ combinations of discretization models ($12$) and EFT-inspired quark-mass-dependence parameterizations ($7$), and combined them with the AIC. At the physical point, we found the $\rho$ pole at
\begin{align*}
    \sqrt{s}_\rho = 744(11)(7) - i\,69(7)(3)~\text{MeV},
\end{align*}
where the first uncertainty is the AIC-weighted fit uncertainty, which includes the statistical and energy-extraction systematic uncertainties, and the second is determined by the variation of the central values over the $84$ parameterizations; this variation arises mainly from the fits with four-parameter mass dependence, $p_5$--$p_7$. At the present precision, the discretization effects are not statistically resolved. In none of the $84$ fits does a discretization coefficient differ from zero by more than $2\sigma$. 
In contrast, the pion-mass-dependence parameters of $m_\rho$ in $p_{1,2,3,4}$ are nonzero at about $3.5\sigma$, and the $m_{\eta_s}$-dependence parameters are nonzero at $5$--$7\sigma$ and have larger values than the $m_\pi$-dependence parameters in every fit. This is surprising, given that the strange quark enters only as a sea quark. Since only \texttt{E5} has a significantly deviating strange-quark mass, we cannot reliably conclude that a large coefficient of the $m_{\eta_s}$-dependence parameter is a physical effect of the change in the strange-quark mass.

The pole trajectory shows the expected behavior of a $P$-wave resonance whose decay phase space opens as the pion mass decreases. At the physical strange-quark mass, the pole moves from about $790 - i\,17$~MeV at $m_\pi \approx 317$~MeV to the physical-point value, and the width grows by roughly a factor of four. The three-parameter mass-dependence models give monotonic trajectories, while the four-parameter forms can give non-monotonic ``boomerang'' trajectories.

The continuum, physical-point phase shift curve mostly agrees with the experimental data of refs.~\cite{Estabrooks:1974vu, Protopopescu:1973sh}. Our width agrees at the $1.5\sigma$ level with the PDG $T$-matrix pole range, $(761\text{--}765) - i\,(71\text{--}74)$~MeV~\cite{ParticleDataGroup:2026mpi}. Other lattice calculations at or extrapolated to the physical pion mass report Breit--Wigner parameters rather than pole positions: $m_{\rho}^{\rm BW}=786(20)$ and $\Gamma_{\rho}^{\rm BW}=180(6)$~MeV for $N_f=2$~\cite{Fischer:2020yvw}; $m_{\rho}^{\rm BW}=796(5)(15)(48)(2)$ and $\Gamma_{\rho}^{\rm BW}=192(10)(28)(12)(0)$~MeV for $N_f=2+1$ at physical quark masses and a single lattice spacing~\cite{Boyle:2024grr}; and $m_{\rho}^{\rm BW}=781.6(10.0)$ and $\Gamma_{\rho}^{\rm BW}=146.5(9.9)$~MeV from a continuum and chiral extrapolation of nine $N_f=2+1$ ensembles~\cite{Wang:2025hew}. 

Our approach's precision is currently limited by the number of ensembles, as four ensembles allow at most four quark-mass parameters per $K$-matrix parameter. A second ensemble with a strange-quark mass away from its physical value, and a finer ensemble near the physical pion mass, would separate the strange-quark and lattice-spacing dependence more cleanly. The continuum amplitude obtained here, with its explicit quark-mass dependence, is an input to extending our study of the $B\to\rho(\to\pi\pi)\ell\bar\nu$ transition \cite{Leskovec:2025gsw} to the physical point.

\newpage

\appendix
\section{Polynomial $K$-matrix terms and $P$- and $F$-wave mixing} \label{ap:amps}
The available energies probe only the region below the higher-lying resonances, namely $\rho(1450)$, $\rho(1700)$, and $\rho_3(1690)$. We therefore use low-order polynomials in the reduced $K$ matrices (cf section \ref{ssec:QC}) to test sensitivity to smooth contributions in this region:
\begin{align}
    \tilde{K}_P = \frac{g^2}{m^2_\rho -s} + \sum_{n=0}^{n_{\rm max}} c_n s^n.
\end{align}

\begin{figure}[htbp]
\centering
\includegraphics[width=\textwidth]{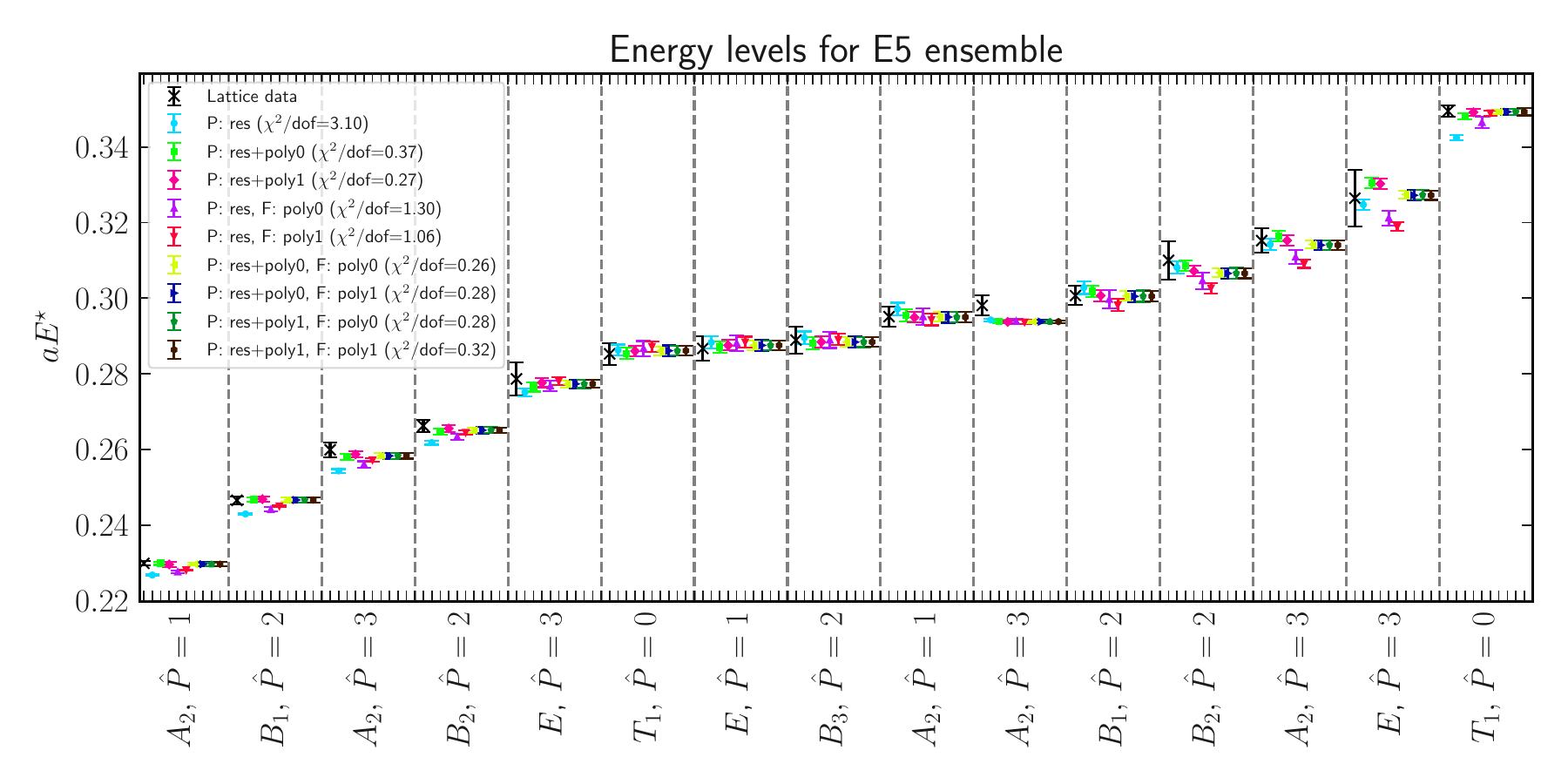}
\caption{Comparison of modeled energies to lattice energies on the \texttt{E5} ensemble, for the different models. The legend lists the $\chi^2/\text{dof}$ for each model. The notation $P$ and $F$ denote the partial waves $l=1$ and $l=3$, respectively; \texttt{res} stands for the resonant term with a pole, \texttt{poly0} denotes the inclusion of the 0th-order polynomial expansion, and \texttt{poly1} the 1st-order polynomial expansion in the chosen partial wave.\label{fig:pfE5}}
\end{figure}

\begin{figure}[htbp]
\centering
\includegraphics[width=\textwidth]{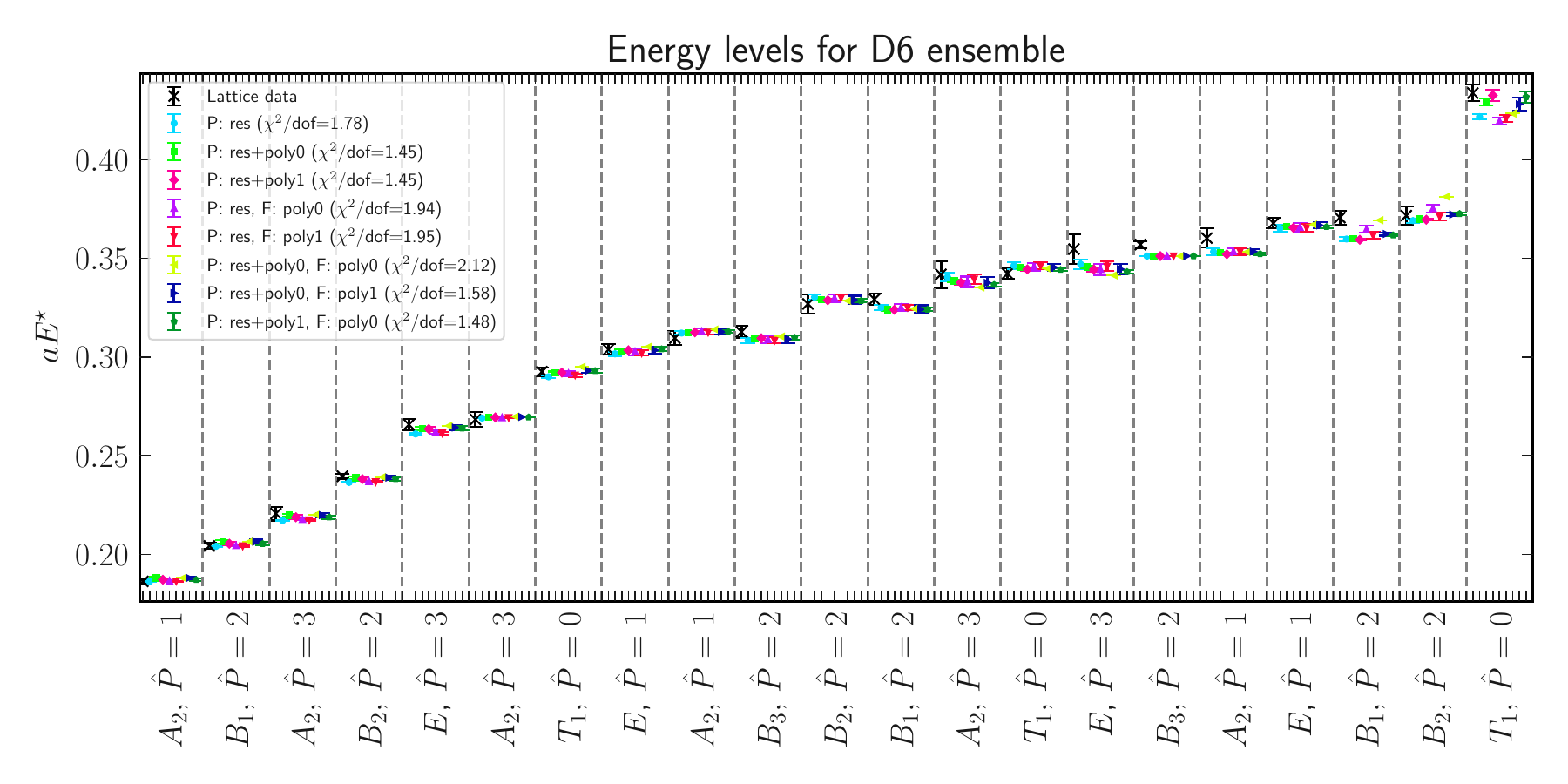}
\caption{Like figure \protect\ref{fig:pfE5}, but for the \texttt{D6} ensemble. \label{fig:pfD6}}
\end{figure}

To probe for the $\rho_3(1690)$, we included the $F$-wave  $(l=3)$ in the quantization-condition fit. For the irreducible representation $\hat{P} =0, \Lambda = T_1$, we use the formalism in \cite{Luu:2011ep}, rewritten in a $T$-matrix format to match equation (\ref{eq:QC}). For the quantization condition in the boosted irreducible representations, we project the $F$ and $T$ matrices onto irreducible representations analogously to \cite{Luu:2011ep}. In the $F$-wave, we only include the polynomial terms:
\begin{align}
    \tilde{K}_F = \sum_{m=0}^{m_{\rm max}} c_m s^m.
\end{align}
We fit the combined amplitude separately for each ensemble with polynomial expansions up to $n_{\rm max}, m_{\rm max} < 2$. The fits are shown in figures \ref{fig:pfE5} and \ref{fig:pfD6}. 

For the \texttt{E5} ensemble, the fit without any polynomial parameters shows large discrepancies between the data (black crosses) and the model prediction (light blue dots) at both low and high energies, as reflected in a high $\chi^2/\text{dof} = 3.1$. The inclusion of a constant polynomial term to $l=1$ (light green square) significantly improves the fit quality, reducing $\chi^2/\text{dof}$ to $0.37$. Including both constant and linear terms (pink diamond) has diminishing returns, as $\chi^2/\text{dof}$ is already very low. The polynomial  expansion in $l=3$ also improves the fit quality with $\chi^2/\text{dof} = 1.3$ for zeroth order (purple up triangles) and $\chi^2/\text{dof} = 1.1$ for a polynomial containing constant and linear terms (red down triangles). Fitting combinations of the $l=1$ and $l=3$ polynomial expansions again has diminishing returns. 

On the \texttt{D6} ensemble, the $l=1$ model without any polynomial terms (light blue dots in figure \ref{fig:pfD6}) deviates from the data (black crosses) in the high-energy region, with $\chi^2/\text{dof} = 1.8$. However, the addition of polynomial terms does not sufficiently improve our fit quality, with the best fit including a 0th-order expansion in $l=1$ and $\chi^2/\text{dof} = 1.45$ (light green squares). Including the polynomial terms in $l=3$ has no effect and even worsens the $\chi^2/\text{dof}$, as seen in figure \ref{fig:pfD6}. 

The addition of polynomial expansions fails to adequately describe the energy spectrum of the \texttt{D6} ensemble. As we were unable to model the effects in the upper part of our energy spectrum, we decided to remove five energies listed in the section \ref{sec:Espec}. These energies lie high in the spectrum, near the $K\bar{K}$ threshold, and may not be extracted reliably due to the small correlation matrices. The two removed energies on the \texttt{D6} ensemble are also near the lowest $4\pi$ states. With our interpolator basis, we cannot exclude that we see a linear combination of a two-pion and a four-pion state. 

At the same time, we removed four energies that were very close to the noninteracting levels, as we described in detail in section \ref{sec:Espec}. Fitting the resonance in $l=1$ without any polynomial terms gives us $\chi^2/\text{dof} = 1.1$ for the \texttt{D6} ensemble with four fewer energies and $\chi^2/\text{dof} = 0.7$ for the \texttt{E5} ensemble with three fewer energies.

\acknowledgments
    We thank Marcus Petschlies for collaboration on related earlier work. We thank the authors of ref.~\cite{Yoo:2026uul}, as well as Will Detmold, David Murphy, and Phiala Shanahan for providing the gauge-field configurations. We are grateful to Daniel Mohler for discussing $l=3$ contributions. Computations for this work were carried out in part on facilities of (1) the USQCD Collaboration, which are funded by the Office of Science of the U.S.~Department of Energy, (2) the Leibniz Supercomputing Centre under project pr27yo, which is funded by the Gauss Centre for Supercomputing,  (3) the National Energy Research Scientific Computing Center, a Department of Energy User Facility, (4) the Extreme Science and Engineering Discovery Environment (XSEDE), which was supported by National Science Foundation grant number ACI-1548562, and (5) the Oak Ridge Leadership Computing Facility, which is a DOE Office of Science User Facility supported under Contract DE-AC05-00OR22725, (6) Blue Waters, which is supported by the National Science Foundation (awards OCI-0725070 and ACI-1238993), (7) Los Alamos National Laboratory, (8) the HPC system Vega funded by the HPC RIVR consortium and EuroHPC JU. L.L. acknowledges the project was financially supported by the Slovenian Research Agency through projects P1-0035, J1-70036 and N1-0360. S.M.~is supported by the U.S. Department of Energy, Office of Science, Office of High Energy Physics under Award Number DE-SC0009913. J.N. acknowledges support by the U.S. Department of Energy, Office of Science, Office of Nuclear Physics under grant DE-SC-0011090. A.P.~acknowledges support by the ``Fundamental nuclear physics at the exascale and beyond'' project under grant DE-SC0023116. S.P. acknowledges partial support from the projects IMAGE-N (EXCELLENCE/0524/0459), LENA-QCD (POST-DOC/0925 - ROandOther/0180) and partonWF (VISION ERC/0525/0010) funded by the European Regional Development Fund and the Republic of Cyprus through the Cyprus Research and Innovation Foundation.

\bibliographystyle{JHEP}
\bibliography{main.bib}

\end{document}